\documentclass[final]{svmult}

\usepackage[LGR, T1]{fontenc}
\usepackage[bottom]{footmisc}
\usepackage{graphicx,newtxtext,newtxmath,csquotes,enumerate,amsfonts,latexsym}

\usepackage[backend=biber,style=biblatex-spbasic,sorting=nyt,natbib=true,uniquename=init,dashed=false]{biblatex}
\bibliography{ArenhartArroyoPreprint.bib}

\begin{document}
\mainmatter

\title*{The roads to non-individuals}
\subtitle{(and how not to read their maps)}
\author{Jonas R. B. Arenhart \and Raoni Arroyo}
\authorrunning{Jonas R. B. Arenhart and Raoni Arroyo}
\institute{Jonas R. Becker Arenhart \at Department of Philosophy, Federal University of Santa Catarina, Florianópolis, SC, Brazil. \at Graduate Program in Philosophy, Federal University of Maranhão, São Luís, MA, Brazil. Partially funded by CNPq, \email{jonas.becker2@gmail.com}
\and Raoni Wohnrath Arroyo \at Department of Philosophy, Communication, and Performing Arts, Roma Tre University, Rome, Italy. Support: grant \#2022/15992-8, São Paulo Research Foundation (FAPESP). \at Centre for Logic, Epistemology and the History of Science, University of Campinas, Campinas, Brazil, \email{rwarroyo@unicamp.br}}

\motto{Forthcoming in J. R. B. Arenhart, R. W. Arroyo (eds.), Non-Reflexive Logics,
Non-Individuals, and the Philosophy of Quantum Mechanics: Essays in Honour of the Philosophy
of Décio Krause, Springer, Synthese Library 476,
\url{https://doi.org/10.1007/978-3-031-31840-5_5}.}

\maketitle\label{chap:arenhart-arroyo}
\begin{refsection} 
\abstract{Ever since its beginnings, standard quantum mechanics has been associated with a metaphysical view according to which the theory deals with non-individual objects, i.e., objects deprived of individuality in some sense of the term. We shall examine the grounds of the claim according to which quantum mechanics is so closely connected with a metaphysics of non-individuals. In particular, we discuss the attempts to learn the required `metaphysical lessons' required by quantum mechanics coming from four distinct roads: from the formalism of the theory, treating separately the case of the physics and the underlying logic; from the ontology of the theory, understood as the furniture of the world according to the theory; and, at last, we analyze whether a metaphysics of non-individuals is indispensable from a purely metaphysical point of view. We argue that neither non-individuality nor individuality is not to be found imposed on us in any of these levels so that it should be seen as a metaphysical addition to the theory, rather than as a lesson from it.}

\section{Introduction}
\label{arenhart-arroyo-sec:1}

Ever since its beginnings, standard quantum mechanics has been associated with a metaphysical view according to which the theory deals with \textit{non-individual objects}, i.e., objects deprived of individuality in some sense of the term. However, it is also a well-learned lesson of current metaphysical methodology that metaphysical terrain is slowly and hardly won, if ever. That is, one cannot claim to have learned metaphysical lessons so easily from quantum mechanics, because the theory does not wear its metaphysics on its sleeves. In this sense, if quantum mechanics could give us a helping hand, at least in determining a metaphysics of non-individuality to its entities, it would have been such a great methodological victory for metaphysics.

But can we learn such lessons? In this paper, we investigate precisely this question and check for the source of non-individuality and its justification. In more precise terms, we inquire into what kind of justification could be attributed to the claim that quantum entities are non-individuals. Does this follow from quantum mechanics itself? Is it attributed to an external source other than physics? If so, does this source grant any kind of justification for such an attribution, or is it floating free from the relevant physics? 

The question of the justification for such attribution of metaphysics is important not only for epistemic reasons, related to the methodological question of the metaphysics of quantum mechanics, but also, we believe because it highlights important features of non-individuality that are not yet discussed in the literature, and which are related with the very characterization of non-individuality. To claim that quantum entities are \textit{non}-individuals, notice, is to advance a characterization in negative terms: quantum entities have \textit{lost} a feature that classical entities do have, individuality. This view is called `the Received View' on quantum non-individuality, given that it is the view that we received as the standard account of quantum entities by the likes of Schr\"{o}dinger and Weyl \citep[see][chap.3]{frenchkrause2006}. The former, for instance, suggested that quantum entities are radically different from their classical counterparts, and dedicated more than one occasion to emphasize this:

\begin{quote}
    This essay deals with the elementary particle, more particularly with a certain feature that this concept has acquired --- or rather lost --- in quantum mechanics. I mean this: that the elementary particle is not an individual; it cannot be identified, it lacks ``sameness''. \textelp{} In technical language it is covered by saying that the particles ``obey'' a new-fangled statistics, either Bose-Einstein or Fermi-Dirac statistics. The implication, far from obvious, is that the unsuspected epithet ``this'' is not quite properly applicable to, say, an electron, except with caution, in a restricted sense, and sometimes not at all. \citep[p.~197]{schrodinger1998}
\end{quote}

Here we have the whole package associated with the Received View (the RV, for short, from now on) in a nutshell. It is claimed that quantum particles have lost their individuality (a metaphysical feature), and that the main reason for that is the presence of new statistics in quantum mechanics (a physical feature), and that the particles lack `sameness' (if that is understood as a failure of the identity relation, then, this is a logical feature). There are distinct features that a non-individual have, and they are related somehow. As a result, we shall discuss whether these features help us characterize non-individuality and whether they do help us derive any kind of justification for non-individuality from quantum mechanics. 

Before we present the plan for this paper, however, let us advance a distinction that makes for the background of this paper. We are assuming a distinction between ontology and metaphysics, where the former is understood as a kind of subset of the latter; this is rather traditional but is not undisputed these days, so we state it here as a starting point, given that it is not our aim to defend the correction or fruitfulness of the distinction. In a nutshell, `ontology' deals with questions of existence, and metaphysics with questions of nature \citep[see][]{arenhart2012, arenhart2019, arenhartarroyo2021manu, hofweber2016, thomsonjones2017}. In this way, we understand that the ontological questions are: \textit{what exists, and how are those entities that exist?} `Metaphysics' would go further, speculating about the \textit{nature} of what was obtained in ontology, that is, developing from the grounds provided by an ontology. 

The first stage of ontology, in the more naturalistic context we concentrate on here, would be to form a catalog about what exists according to a given scientific theory.\footnote{This particular task of ontology is named in several other ways in the literature, many of them metaphorical, such as the establishment of an inventory of the furniture of the world according to the theory in question. Note that at this stage we are also assuming a metaontological posture inspired by the Quinean tradition, according to which we can speak about the ontological catalog associated with certain theories, although we do not wish to take sides here on questions of whether existence is better represented by quantification or predication, and the like.} To refer to the existents of a theory, we choose the word `entity', because it seems to be more neutral than `thing', `stuff', or `object'.\footnote{This type of caution has gained attention; we see this, for example, in \citet[p.~2]{meincke2020}: ``I deliberately speak of `entities' rather than (as many scholars do) of `objects' to leave open the possibility that also non-object-like entities, such as processes, events, states of affairs, structures etc''. We shall return on this matter, albeit briefly, in section \ref{arenhart-arroyo-sec:4}.} As our discussion will focus on quantum mechanics, it seems safe to say that in a straightforward reading, the theory tells us that there are electrons, protons, and other entities typically conceived of as particles. They are part of the furniture of the world according to quantum mechanics. Electrons exist according to the theory, but we still don't know what kind of entities they are; at least, quantum mechanics doesn't give us straightforwardly that kind of information. So, the second task of ontology would be to organize these entities that exist in more general categories, types of entities. In this paper, given our more specific interests, we will focus on the category of `objects' (other categories include properties and relations, and events). Suppose electrons are objects: what does that say about the accompanying metaphysics? Well, we can understand them metaphysically at least in two different ways, e.g., as individuals or as non-individuals. But what are individuals? And what are non-individuals? To answer these kinds of questions is to attribute what we call a `metaphysical profile' to quantum entities, or, what we take to be equivalent here, to describe their `nature'.

Given this distinction, recall, our problem is to discuss the notion of non-individuality and the kind of justification one may have to say that quantum mechanics provides for a metaphysics of non-individuals. We are concerned with the source of non-individuality, which would also confer to the notion a kind of legitimacy. As we have briefly discussed already, there are distinct features that are used to characterize a non-individual, metaphysical, logical, and physical ones. Could they shed any light on the justification for the attribution of non-individuality for quantum mechanics? Our goal is to investigate precisely that.  

In section \ref{arenhart-arroyo-sec:2}, we will try to answer this question by attributing this role to logic; in section \ref{arenhart-arroyo-sec:3}, by seeing physics as in charge for the answer; in section \ref{arenhart-arroyo-sec:4}, ontology; finally, in section \ref{arenhart-arroyo-sec:5}, we will look for the metaphysical profile of non-individuals in metaphysics. We will go through these different areas of knowledge with a \textit{Wanted} sign, checking what kind of information on non-individuality one may obtain from them, and conclude in section \ref{arenhart-arroyo-sec:6}.

\section{From logic to non-individuality}
\label{arenhart-arroyo-sec:2}

We start with attempts to find non-individuality in logic. The plan, in a nutshell, is: if non-individuality could be codified logically, in terms of the underlying logic, then, perhaps, one could have some kind of justification for a clear notion of non-individual in a system of logic incorporating this notion. That is, perhaps non-individuality is a matter of logic, and we could justify non-individuality if we could justify the adoption of such a logic. 

Recall now Schrödinger's quote above. Part of the idea of non-individuality comes from the idea that quantum entities lack `sameness'. One typical way, which has been traditionally wedded to the Received View in such a way that it is currently conflated with the RV itself, is to follow one of the many indications by Schrödinger in the quote presented, that quantum particles lack ``sameness'' (see the discussion in \cite{arenhart2017synt}). This has been taken literally as an inspiration for the development of non-classical logics codifying this suggestion, that is, systems of logic in which the idea of identity loses its meaning due to lack of application; these systems have been baptized \textit{non-reflexive logics}, due to the failure of the reflexive law of identity (see \cite[chap.~7-8]{frenchkrause2006} for the standard presentation of the systems).\footnote{However, non-reflexivity indicates only part of what is failing here; the system does not express any result concerning identity for the intended quantum entities, so every property of identity fails.} The intuition behind this formulation of the RV is that \textit{non-individuals are entities to which the logical relation of identity does not apply}. As non-reflexive logics capture this notion of non-individuals, the expectation is to find the metaphysical profile of non-individuals in non-reflexive logics. As \citet[p.~22]{french2019} has put it, referring to quasi-set theory (a form non-reflexive set theory), ``forms of non-standard set theory have been presented as a framework in which to understand the claim that they are non-individuals.'' Also, about the relation between non-reflexive logics and non-individuality:
\begin{quote}
    These developments [of non-reflexive logics] supply the beginnings of a categorial framework for quantum `non-individuality' which, it is claimed, helps to articulate this notion and, bluntly, make it philosophically respectable. \citep[sect.~5]{sep-qt-idind}
\end{quote}

So, the logical framework is supposed to make a substantial job in delivering at least the general contours of the metaphysics of non-individuals, its `articulation'. However, notice that a system of logic, per se, does not have any commitment to a metaphysical picture of reality unless we attribute it one (for the case of the RV and non-reflexive systems, see \cite{arenhart2018}). There must be something like an intended interpretation of the lack of identity as meaning specifically failure of individuality, otherwise, no lesson is learned from that system. Standard classical logic without identity could also be seen as a non-reflexive logic, although it is not typically seen as articulating a metaphysics of non-individuals. So, there must be more to it if logic is going to teach us something about non-individuality. 

Of course, one way to establish such an intended meaning, from which metaphysical lessons could be drawn, may be obtained if we add a metaphysical reading over the system, a reading that is not necessarily there but is the intended application of the system. Perhaps this is what is actually done. By recalling that individuality may be understood in terms of haecceities, non-qualitative properties that items have, and which intuitively mean `being identical to itself',\footnote{So, Plato's haecceity is the property `being identical to Plato'.} \citet[pp.~13--14]{frenchkrause2006} indicate that the metaphysical meaning of non-individuality may be understood in terms of the lack of a haecceity, and the formal expression of that is the failure of identity:
\begin{quote}
\textelp{} the idea is apparently simple: regarded in haecceistic terms, ``Transcendental Individuality'' can be understood as the identity of an object with itself; that is, `$a = a$'. We shall then defend the claim that the notion of non-individuality can be captured in the quantum context by formal systems in which self-identity is not always well-defined, so that the reflexive law of identity, namely, $\forall x(x = x)$, is not valid in general. \citep[pp.~13--14]{frenchkrause2006}
\end{quote}

They recognize that by establishing such an association they ``are supposing a strong relationship between individuality and identity \textelp{} for we have characterized `non-individuals' as those entities for which the relation of self-identity $a=a$ does not make sense'' \citep[p.~248]{frenchkrause2006}. This association, however, clearly does not come from logic, but from specific choices of a particular characterization of individuality that relates it to identity and, consequently, a relation of lack of identity to lack of individuality.

So, what is really going on here is that we have a different attitude towards non-individuality and its relation to the formalism of non-reflexive logics than the one we expected: it is not that the formal system teaches us about non-individuality, but rather, there is a previous notion of non-individuality that the system is supposed to capture. In this sense, the metaphysics of non-individuality comes from \textit{outside} of the formal system and is neither derived from it nor articulated by it. The dignity of the Received View, in this case, or its philosophical respectability, does not rest on the success of the distinct systems of non-reflexive logics, but rather on the coherence of the idea that non-individuality may be reasonably couched in terms of the lack of haecceity. 

However, this is now a metaphysical view that is somehow \textit{described by} non-reflexive logics, not \textit{extracted from} such logics. There are no special lessons learned for this metaphysics that derive from the non-reflexive systems, rather, the expectation is that such systems must correctly capture the basic tenets of lack of haecceity. In this sense, the explanatory direction goes from metaphysics to logic, one must have the metaphysics clear beforehand, to develop the logic that will describe it and not the other way around. 

Things get more complicated for the non-reflexive approach to the Received View when one considers that such logic is not even necessary for the Received View. If one concedes that quantum entities are non-individuals but refuses to follow the suggestion of Schrödinger and understand it in terms of lack of identity, and more, refuses to follow French and Krause couching lack of individuality in terms of lack of haecceity, one can still adopt a form of non-individuality. Indeed, one may adopt another suggestion, briefly advanced but not developed by \citet[chap.~4]{frenchkrause2006}, for instance, and see individuality as conferred by the spatiotemporal location. In this case, given that quantum theory is typically seen in most of its formulations as suggesting that particles do not have a well-defined location all of the time, then, individuality fails. However, notice that this approach is not formulated in terms of the lack of identity. This illustrates the general thesis that, if non-individuality is to be understood in terms of a lack of individuality, then, our metaphysical framing of the notion will depend on a theory of what individuals are, to begin with, and distinct such theories will give rise to distinct theories of non-individuals \citep[see][]{arenhart2017synt}. Most of them will not require that the logical relation of identity is abandoned, and that non-reflexive systems be adopted. Most metaphysics of non-individuals will be quite compatible with classical logic, and it would be incorrect to think that classical logic teaches us anything special about \textit{them}. This corroborates the claim then, that logic, per se, does not teach us metaphysical lessons about non-individuality (and neither about individuality too, of course).

\section{From physics to non-individuality}
\label{arenhart-arroyo-sec:3}

Perhaps we need a more concrete jury, like an empirical science. As the quote in the introduction indicates, one of the main reasons that Schrödinger, among others, had to suggest that quantum mechanics deals with non-individuals comes from the new kinds of statistics that the theory obeys. So let's see if a minimal portion of the formalism of quantum mechanics itself could succeed where logic failed to offer us a metaphysical profile for non-individuals.\footnote{Notice: the idea that logic failed to offer us a metaphysical profile means here that, from logic alone, one can neither derive the claim that quantum entities are non-individuals and nor obtain any specific information on the notion of non-individuality that could be used to describe the nature of quantum entities, were they non-individuals.}

Although much discussion has already been made about the role of statistics in quantum mechanics in deriving metaphysical lessons, the issue that Schrödinger wanted to highlight may be put in quite simple terms, contrasting the new quantum statistics with the classical Maxwell-Boltzmann statistics. The idea that classical particles have individuality, even in the situations in which one considers that they do have all the same properties (that is, when they are \textit{indiscernible}) is famously encapsulated in the way that Maxwell-Boltzmann's statistics work. Let us illustrate it with the case in which two particles, labeled $1$ and $2$, must be distributed in two states, $A$ and $B$, which are equiprobable. We have the following possibilities (where $A(1)$ means that particle $1$ is in state $A$, and so on):
\begin{enumerate}
\item $A(1)A(2)$;
\item $B(1)B(2)$;
\item $A(1)B(2)$;
\item $A(2)B(1)$.
\end{enumerate}

All these possibilities are assigned the same weight, that is, $\frac{1}{4}$, and even though the particles are being considered as indistinguishable by their properties (state-independent and state-dependent alike), permutations of particles are counted as distinct possibilities in the steps 3 and 4; in fact, they differ only by a permutation of the labels of the particles. So, if the particles may be indistinguishable, what accounts for the difference in the two situations described in 3 and 4? Well, it seems that there must be something in the particles that explains the fact that exchanging 3 with 4 makes a difference, and this something is thought to be the particles' individuality. 

The individuality of classical particles may be explained in a variety of ways (and the metaphysics of individuality for classical particles gets underdetermined by classical physics). It is possible to adopt the kind of individuality principles that add some metaphysical ingredient, the ``Transcendental Individuality'' approach, with such ingredient understood in terms of a substratum, or a haecceity\footnote{And they are distinct approaches, given that the substratum is a \textit{particular} item that is added to the composition of the particular object, while haecceity is a non-qualitative \textit{property}.} or one may adopt a bundle theory, individuating the particles by granting that each individual has some unique set of properties that may be used to characterize it, and which accounts for its individuality (see \cite[chap.~1]{frenchkrause2006} for the basic approaches to individuality). The latter can always obtain in classical mechanics by the adoption of spatial position as a property, and a Principle of Impenetrability, which holds in classical mechanics, and which says that no two particles occupy the same position. 

In quantum statistics, on the other hand, permutations of indistinguishable particles are not counted as giving rise to a distinct state. This fact is related to the already mentioned Permutation Symmetry in quantum mechanics, and is behind the typical allegations of loss of identity and individuality; that is, quantum mechanics itself would be the source of the failure of individuality. This is illustrated as follows. For two systems, labeled $1$ and $2$, distributed in two possible states $A$ and $B$, we can have the following possibilities:
\begin{enumerate}
\item $A(1)A(2)$;
\item $B(1)B(2)$;
\item $\frac{1}{\sqrt{2}}(A(1)B(2) \pm A(2)B(1))$.
\end{enumerate}

In fact, we have two different kinds of statistics here: Bose-Einstein (BE) for bosons and Fermi-Dirac (FD) for fermions. The difference comes in the third possibility, because bosons have the ``$+$'' sign, and fermions have the ``$-$'' sign. Also, for fermions only the third case obtains; they cannot be distributed according to the first two cases for they cannot be in the same state, due to the Pauli Exclusion Principle, which holds for fermions (but not for bosons). This third possibility indicates, briefly, that the description of a state in which the particles $1$ and $2$ are distributed over $A$ and $B$ must take into account, simultaneously, both cases of possible distributions; notice that swapping the labels of particles does not change the state for bosons, while it just changes the sign of the state for fermions (and does not change the probabilities associated with the state). Permutations of particles are not observable. 

Does this indicate anything special about metaphysics? As we have seen, Schrödinger, and many others, thought so (see \cite[chap.~3]{frenchkrause2006}). The only explanation for this possibility of permuting the particles without resulting in any different state, according to such a view, is that there is nothing in the particles themselves that allow for such a difference that would be had in the states such as $A(1)B(2)$ or $A(2)B(1)$, which would distinguish the particles. So, the particles are not individuals, because they lack precisely that which makes the classical particles individuals. Furthermore, given that permutations are not observable, the idea seems to be that one cannot identify particles, they are not things that one can trace and find in distinct situations. So, these features all suggested to Schrödinger that identity, in general, fails to apply to such particles, giving rise to the now traditional association between non-individuality and lack of identity \citep[see the discussion in][]{arenhart2017synt}. 

The problem with this view is that the association between the behavior of quantum particles as described by the statistics is not so straightforwardly connected with the failure of identity, and also, it is not so straightforwardly associated with non-individuality. Non-individuality, as a failure of identity, does not drop out of the quantum apparatus. Let us check briefly. 

We have already commented in the previous section that the identification of non-individuality and lack of identity is a kind of historical contingency. The lack of individuality, later it was discovered, could be described in many distinct ways \citep[see, in particular, ][]{arenhart2017synt}. Quantum mechanics does not explicitly indicate that it is the particular approach to non-individuality as a failure of identity that it endorses.  

The same could be said about non-individuality itself. As argued at length in \citet[chap.~4]{frenchkrause2006}, quantum theory is compatible with a metaphysics of individuals. It is enough that individuality is attributed to some principle that allows indiscernible objects, like substrata or haecceities. The fact that quantum particles cannot be distinguished by quantum mechanical means does not produce any difficulty for these approaches, whose main motivation is in fact to deal with scenarios in which objects are not discernible and, hence, not able to be individuated by their properties or relations. So, no wonder that quantum mechanics is compatible with a metaphysics of individuals too. The theory says nothing about its metaphysics of individuality and non-individuality. As French has put it, concerning classical mechanics (but the lesson applies to quantum mechanics too)\footnote{The same point was made already in \citet[pp.~344--345]{arenhart2012} in connection to quantum mechanics.}:

\begin{quote}
\textelp{} that metaphysics does not supply the kinds of `objects' over which appropriately formalised versions of scientific theories could be said to quantify (do first order formulations of classical statistical
mechanics quantify over \textit{haecceities}, for example?!). \citep[p.~216, original emphasis]{french2018jgps}
\end{quote}

That is, metaphysics is not in physics. We cannot hope to get from the theory the metaphysical descriptions that would be required to confirm that the entities dealt with in quantum mechanics are non-individuals, and even for those that believe that quantum mechanics deals with non-individuals, the particular approach to non-individuality is also not determined by the theory. Perhaps it is in another place that we should look for these metaphysical lessons. 

\section{From ontology to non-individuality}
\label{arenhart-arroyo-sec:4}

In some sense, ontology is a \textit{naturalizable} philosophical discipline, i.e., its major question can be answered by science --- at least in part, as we shall see. In particular, by physics. So \textit{maybe} ontology can tell us what non-individuals are. Let's take a closer look at this claim. As \citet[p.~222]{esfeld2019} claims, a physical theory is responsible for offering us a dynamic law and an ontology, i.e., to spell out ``[w]hat is the law that describes the individual processes that occur in nature (dynamics) and what are the entities that make up these individual processes (ontology)''.

In this sense, as \citet[p.~2]{durrlazarovici2020} recognize, ``[t]he ontology of a physical theory specifies what the theory is about''. Furthermore, in quantum mechanics, as \citet[p.~298]{ruetsche2018} acknowledges, this is largely dependent on the interpretation adopted: hence, ``\textelp{} an interpretation of QM tells the realist about QM what she believes when she believes QM''. As this is pretty standard, we recall the current state of quantum foundations very briefly as follows. An interpretation of quantum mechanics is a solution to the measurement problem, whose standard way --- i.e., adopting Maudlin's (\citeyear{maudlin1995}) taxonomy --- is stated as a checklist of three simple questions about the quantum-mechanical (e.g. the wave function) description: i) is it complete? ii) is it linear? iii) does it yield definite (unique) measurement outcomes?\footnote{A proof of inconsistency may be found elsewhere \citep[see][]{maudlin1995, esfeld2019}.} The most elementary definition of an interpretation of quantum mechanics, that is, of a solution to the aforementioned measurement problem, is by the negative answer to one of the questions, which poses a trilemma \citep[see][p.~47]{durrlazarovici2020}. 

To deny the first implies accepting theories of hidden variables, such as Bohmian mechanics; a negative answer to the second question implies the adoption of collapsed theories, such as the standard quantum mechanics with collapse and the GRW theory; finally, a negative answer to the third question implies the acceptance of an Everettian quantum mechanics, such as the many-worlds interpretation. Concerning ontology, each of these interpretations populates the world with different kinds of entities: Bohmian mechanics postulates the existence of a `pilot wave', which governs the behavior of quantum particles (which exist along with the wave); some collapse interpretations postulate the existence of a `consciousness' endowed with causal powers, separated from matter; some Everettian quantum mechanics postulate the existence of a plurality of `worlds', and so on.

The debate over (non-)individuality is not about an interpretation of quantum mechanics in the sense of offering an answer to the measurement problem; however, it seems evident that it is about `interpreting' quantum mechanics in some sense \citep[see][]{krause-arenhart-bueno2022}. Therefore, a peculiarity of this debate is its dependence on the interpretation adopted. Traditionally, metaphysical underdetermination between individuals and non-individuals is discussed in interpretations that admit the dynamics of collapse \citep[see][]{frenchkrause2006,krausearenhart2016}, but it might be also discussed concerning Bohmian mechanics \citep[see][]{pylkkanen-etal2016} and Everettian quantum mechanics \citep[see][]{conroy2016}.

So far we are working with a very standard notion of the term `ontology' which, according to the Quinean tradition, asks ``\textelp{} what, according to that theory, there is'' \citep[p.~65]{quine1951}. So, in the (fairly) uncontroversial part, quantum mechanics tells us that there are bosons and fermions, among other entities. It is in this sense that we say that ontology has been `naturalized': after all, we can extract these entities from the theory, it commits itself ontologically to it \citep[see also][]{arenhartarroyo2021manu}. As we have seen, other entities can be extracted depending on the interpretation adopted. In this way, a catalog or inventory is formed about what exists according to each interpretation. This inventory does not answer as to whether the theory deals with non-individuals, and what they are.

In order to properly address this question of non-individuality through ontology, it could be helpful to introduce another aspect of ontology. The aspect of ontology we have in mind is explicitly acknowledged, for instance, by \citet[p.~30]{ney2014}, when she describes ontology as dealing not only with the issue of the furniture of the world but also with ``a particular theory about the types of entities there are''. That is, we need to establish an ontological \textit{catalog of reality}, listing what things exist (tables, chairs, electrons, Everettian worlds, Bohmian pilot waves, etc.); but we also need to establish \textit{ontological categories}, classifying into \textit{types} the entities of the catalog (examples of types: objects, immanent powers, structures, wave functions, etc.).

While interpretations are clearly concerned with the catalog, and one could, at least in principle, attempt to find in the theory along with an interpretation the answer to the catalog question, the same cannot be said about the \textit{types} of entities. In fact, the theory does not say anything explicitly about the type of entities it deals with, and this shall cast doubt on the project to attempt to extract any information about non-individuality from the ontology of quantum mechanics. Let us check.

One could hope that ontology and quantum theory have a close connection. It could go along the following lines: quantum theory commits us with objects (ontological type) behaving in many respects quite different from what individual objects typically do (the nature of such objects). These would be the first steps in framing a connection between ontology and the nature of non-individuals. It would enjoy the benefit of epistemic authority derived from quantum mechanics itself. It is common to find in the literature that the focus on ontological aspects is a metametaphysical attitude typical of metaphysical naturalism \citep[see][]{French2018RealMetaph, arroyoarenhart2019}. That is, naturalists maintain ontology as the privileged part of metaphysics, often reducing it to one another. After all, ontology can be obtained from science. Thus, some naturalists, like Maudlin, conclude that:

\begin{quote}
    Metaphysics is ontology. Ontology is the most generic study of what exists. Evidence for what exists, at least in the physical world, is provided solely by empirical research. Hence the proper object of most metaphysics is the careful analysis of our best scientific theories (and especially of fundamental physical theories) with the goal of determining what they imply about the constitution of the physical world. \citep[p.~104]{maudlin2007}
\end{quote}

Realists of this type, according to Magnus' (\citeyear{magnus2012}; see also \cite{French2018RealMetaph}) characterization, are `shallow' scientific realists. That is, scientific realists who feel epistemically safe going only as far as science goes --- no further steps. And science goes --- they say --- only to the shallow waters of ontology. Again, some suggest that ontological proposals are epistemically justified because they are somehow extracted from science. The problem is that people with very different ideas about ``lessons learned from quantum mechanics'' seem to have the same belief. As a sample, we can showcase three alternatives to object-based ontology:

\begin{enumerate}
    \item \textit{Ontic structural realism} (OSR): ``structures'' are the basic elements of ontology \citep[see][]{french2014};

    \item The \textit{logos approach}: ``immanent powers'' are the basic ontological stuff \citep[see][]{derondemassri2021};

    \item The \textit{wave function realism}: the ``wave function'' is the basic stuff \citep[see][]{albert2013}.
\end{enumerate}

The curious thing is: how can we `read off' three ontologies so different from the same quantum theory? The lesson seems to be rather than quantum theory teaches us no lesson about it. Perhaps even the hope of extracting the ontology of types of entities, when we are restricted to a shallow realism, is too high. This seems to be impossible. A first argument is the finding that ontological results, of the types of ontology, are underdetermined by physics. But perhaps the most radical aspect is the following: the type-aspect of an ontology associated with interpretations of quantum mechanics, is \textit{placed on} the catalog. Unlike the catalog-aspect, the type-aspect of ontology involves the \textit{creativity} of the metaphysician of science, rather than \textit{extraction}. Thus, this shallow option cannot even come close to answering the profile of non-individuals, given that it does not guarantee even an ontology of objects. One could argue that the shallow realist never set out to argue about the metaphysical profile of entities. However, if that is the answer to the concern with non-individuality, then, what we are indicating is that the shallow realist falls short of the task she proposes: that of extracting the ontology from the theory. 

An alternative that seems more appropriate would be to interpret the development of these ontologies of types of entities as an addition to what the theory says so that the process of `reading off' the ontology would be better understood as a metaphor that indicates a better fit that each author sees, concerning their proposal, with what the theory provides. In this sense, we agree with Chakravartty's diagnosis, when evaluating the OSR venture as offered by French, in these terms:

\begin{quote}
    When French speaks of reading ontology from fundamental physics, what he is doing, in fact, is implicitly appealing to some (one hopes) defensible criterion or criteria which he takes to point toward a \textit{preferred} interpretation of the relevant physics --- an interpretation which is (one hopes) demonstrably superior to others.
    \citep[p.~11, original emphasis]{chakravartty2019}
\end{quote}

To focus on the OSR only, the point is that, as \citet[p.~29]{esfeld2013} states, ``OSR is not an interpretation of QM in addition to many worlds-type interpretations, collapse-type interpretations, or hidden variable-type interpretations'', and this is so because, in addition to the postulate ``radically different proposals for an ontology of QM'', interpretations of quantum mechanics also commit themselves to different dynamics. So even if one takes the `shallow' road to realism, such a road is a slippery slope between a) not having all ontology informed by science, and b) not having an epistemic (objective) guarantee to adopt an alternative to the detriment of another.

We started out wanting a metaphysical profile for non-individuals, and we saw that `shallow' realism, as a naturalistic approach to ontology, is a road that didn't lead us to that. It fails in two important directions: it does not concern itself explicitly with metaphysical profiles, and, worse yet, it does not grant epistemic authority from science to the types-ontology that is advanced in each case. In this sense, there is a failure in granting the required ontology of objects an appropriate justification, and one could see this as more trouble for the non-individuals interpretation. However, this more basic source of trouble is an issue we shall not discuss here. 

\section{From metaphysics to non-individuality}
\label{arenhart-arroyo-sec:5}

An alternative to `shallow' realism, still within the scope of scientific realism, would be the `deep' scientific realism. Here, metaphysicians deliberately venture into the metaphysical waters in search of a metaphysical profile for non-individuals. On the one hand, giving a metaphysical profile seems to be a necessary component of one's scientific realism. After all, according to \citet[p.~26]{chakravartty2007} ``[o]ne cannot fully appreciate what it might mean to be a realist until one has a clear picture of what one is being invited to be a realist about'' (i.e. Chakravartty's Challenge). On the other hand, how to do that? Here's French:

\begin{quote}
    A simple answer would be, through physics which gives us a certain picture of the world as including particles, for example. But is this clear enough? Consider the further, but apparently obvious, question, are these particles individual objects, like chairs, tables, or people are? In answering this question we need to supply, I maintain, or at least allude to, an appropriate metaphysics of individuality and this exemplifies the general claim that in order to obtain Chakravartty’s clear picture and hence obtain an appropriate realist understanding we need to provide an appropriate metaphysics. \citep[p.~48]{french2014}
\end{quote}

So, on what concerns metaphysical issues such as individuality (but not only individuality, of course), physics seems to be the source for the metaphysical profile. This is also indicated in the following claim by French and Krause, where it is thought that one expects that the `theoretical content' of the theory should be articulated in metaphysical terms. That is, the theoretical content and the metaphysics in terms of which it is articulated come from the same source:
\begin{quote}
\textelp{} if that theoretical content is taken to have a metaphysical component, in the sense that the realist's commitment to a particular ontology needs to be articulated in metaphysical terms, and in particular with regard to the individuality or non-individuality of the particles, then the realist appears to face a situation in which there are two, metaphysically inequivalent, approaches between which no choice can be made based on the physics itself. \textelp{} The choice for the realist is stark: either fall into some form of antirealism or drop the aforementioned metaphysical component and adopt an ontologically less problematic position. \citep[p.~244]{frenchkrause2006}
\end{quote}

That is, in the face of metaphysical underdetermination between individuality and non-individuality, one solution is to change the ontological basis, with the hopes that a distinct approach to the general kind of entities that one is assuming the theory deals with could solve the problem of granting a metaphysical description along with the theoretical content (they suggest an ontology of relations in place of an ontology of objects). The great concern here is epistemic since in these are turbulent waters, which no longer find support in scientific theories:

\begin{quote}
    \textelp{} you get only as much metaphysics out of a physical theory as you put in and pulling metaphysical rabbits out of physical hats does indeed involve a certain amount of philosophical sleight of hand. \citep[p.~466]{french1995}
\end{quote}

Hence, the metaphysical profile finds itself `floating free' from physics. Methodologically speaking, we saw that we cannot extract the metaphysical profile from logic, physics, or ontology \citep[see][]{arroyoarenhart2020}; it remains to be seen whether we can extract the metaphysical profile$\dots$ from metaphysics! This is, roughly speaking, the proposal put forward by French and McKenzie, sometimes called the ``Viking approach'' \citep[see][]{french2014}, and sometimes the ``toolbox'' approach \citep[see][]{frenmckenzie2012, frenmckenzie2015, french2018jgps}. The general idea of this approach is as follows: recognizing the need for a metaphysical profile so that we can give content to `deep' scientific realism, metaphysicians would be free to produce their metaphysical theories free from science. These theories, when necessary, can be used for interpretive purposes of science, that is, to provide an adequate metaphysical profile for scientific theories --- thus enabling the adoption of a metaphysically informed (or `deep') scientific realism. From ready-made metaphysical theories --- as could be the case with substance dualism and some interpretations of collapse-based quantum mechanics \citep[see][]{arroyoarenhart2019} --- to methods and strategies available in the metaphysical literature --- such as, for example, methodological lessons that we use to extract from Lewisian modal realism a metaphysical profile to certain interpretations of Everettian quantum mechanics \citep[see][]{wilson2020}.

The major difficulty in attempting to pick metaphysical theories of individuality and non-individuality from metaphysics proper is that the variety of options is just too big, and one ends up losing sight of the epistemic anchorage that quantum mechanics could provide for any such metaphysical adventure. That is, the only restriction that one could find is a straight inconsistency between quantum theory and a given metaphysical package, but this is still not enough to avoid metaphysical underdetermination. Regarding the metaphysical layer, using the Viking approach, we can give a metaphysical body to the ontological bones of each of the interpretations. As there is no anchoring in physics, it is to be expected that there is a metaphysical underdetermination \citep[cf.][]{arroyoarenhart2020}. We are aware that many authors understand metaphysical underdetermination as a motivation for structural realism, and this is the case of the classic case of the individuality of quantum objects \citep{frenchkrause2006}: if our best scientific theories do not provide us with elements to decide between a metaphysics of individuals and a metaphysics of non-individuals, we should modify our ontological basis for structures, where this underdetermination of individuality does not occur \citep[cf.][]{ladyman1998, french2014, french2020motivational}. However, on the one hand, it is not clear what a structure is in metaphysical terms; furthermore, there is no guarantee that, if we have a precise definition of what a structure is in metaphysical terms \citep[cf.][]{arenhartbueno2015}, that definition is unique so that we don't fall prey for the problem of metaphysical underdetermination \citep[cf.][]{french2020whatis}. On the other hand, although structural realism can respond to metaphysical underdetermination, it certainly cannot respond to underdetermination of quantum mechanical interpretations \citep[cf.][]{esfeld2013}, so underdetermination is not a direct guide to the ontology and metaphysics of structures \citep[cf.][]{ruetsche2018}. As we saw in section \ref{arenhart-arroyo-sec:2}, being a non-individual requires something else than values of variables, as \citet[pp.~186--187]{ladyman2016} also acknowledges; the issue becomes even more pressing as there is no consensus about \textit{what} is this ``something else'':

\begin{quote}
\begin{itemize}
\item persistence (French and Redhead (1988))
\item transworld identity
\item countability and determinate identity (Lowe (2003))
\item laws of identity, perhaps including PII
\item absolute discernibility (French and Krause (2006), Muller and Saunders (2008), Caulton and Butterfield (2012))
\item possession of some form of transcendent individuality
\end{itemize}
\citep[pp.~186--187]{ladyman2016}
\end{quote}

As it is to be expected, there is not a single clear picture concerning what individuality is; consequently, as the RV is the position according to which quantum objects \textit{lose} their individuality, it is also unclear what non-individuality is. Worse yet: even if one adopts the idea, following \citet[p.~187]{ladyman2016}, that principles of individuality such as the principle of transcendent individuation are the most fundamental, there are also many options:\footnote{Although some of the terms of the list are sometimes taken as terminological variants of each other, given the lack of a unified definition, it is useful to keep the distinct terms, because they may reflect some differences of usage for some authors.}

\begin{quote}
\begin{itemize}
\item haecceity
\item primitive thisness (Adams (1979))
\item transcendental individuality (Post (1963), French and Redhead (1988))
\item bare particulars
\item individual substances
\item substratum
\item self-individuating elements (Lowe (2003)) \end{itemize}
\citep[p.~187]{ladyman2016}
\end{quote}

On what concerns individuals, as we have seen, the claim that quantum entities are individuals is not itself ruled out, to begin with. What is enough is that the principle of individuality is not based on any notion of qualitative discernibility (these principles would be, at least \textit{prima facie}, incompatible with a quantum mechanical description of objects). One could, for instance, argue that quantum entities are individuated by haecceities, the non-qualitative property of being `identical to itself', and which each individual allegedly bears to itself (see \cite{arenhart2017synt}, \cite[chap.~1]{frenchkrause2006}). Of course, one could complain that this inflates the metaphysics, but when we consider that the Viking merely requires that we look for the appropriate tools in the toolbox, what one finds is that this kind of approach is available, and does some extra job in the trans-world identity (on which we shall not comment). 

Another option would be to confer individuality as a primitive feature of particles. This is the approach taken by \citet{morganti2015}, among others. The plan is not that individuality is conferred by any kind of intrinsic feature, but by a kind of thin notion of identity, which adds nothing else in the constitution of the entity, in metaphysical terms, although at the same time, it plays the role of granting individuality. Here, one may have to choose between the explanatory power of defining individuality in terms of haecceity (which in itself is a bit mysterious), or of conferring it primitively, having to pay the price of losing explanatory power (and this is a reason why \cite[chap.~1]{frenchkrause2006} have restricted their discussion to principles defining individuality, not attributing it primitively). The dispute is common in metaphysics, and it is not something that one could address by appealing to quantum mechanics itself.   

The idea that quantum entities are non-individuals does not suffer a better fate when we consider the toolbox approach. Granted that one needs to provide for a metaphysical profile for non-individuality, that is, the absence of individuality, one could look in the toolbox for diverse principles of individuality and check how they fail in quantum mechanics. This certainly leaves one with plenty of tools available in the toolbox, as the previous discussion makes clear. At first sight, by adopting the association between self-identity and haecceity, one could follow the Received View and attempt to eliminate the latter by eliminating the former. Lack of individuality means lack of haecceity  \citep[see our previous discussion on this issue, and also][]{arenhart2017synt}. However, this is not the only option. As \citet[chap.~4]{frenchkrause2006} have suggested, one could also adopt a view that the principle of individuality consists in the spatiotemporal position of an entity. Given the well-known restrictions imposed on the standard formulation of quantum mechanics, there is not always a well-determined position for a particle, and one could reasonably take it to mean that such particles are non-individuals, they lack the feature defining individuality (curiously, this same principle could be used in Bohmian mechanics as a principle of individuality, so that the toolbox has distinct roles, depending on the ontology, which here is seen as related to the interpretation too).

These examples are enough to show that when one looks for a metaphysics of individuality and also for a metaphysics of non-individuality in the current literature of metaphysics, one does find the tools. The major obstacle is that there are just too many tools, and one cannot expect to get any help \textit{from quantum mechanics} itself on these issues. The plan, of course, was that metaphysics, as a toolbox, could help us get a clearer picture of quantum theory. But in the measure that there are many different metaphysical pictures available, with no direct connection with quantum mechanics, one ends up with no clear picture, but with confusion. 

The situation improves, but not so much if we take a distinct approach to the relation between metaphysics and ontology. One could suggest that a metaphysical profile should be handcrafted, \textit{tailored} for non-individuals. Doing so involves modifying the actual notions of non-individuality in order to make the metaphysical concepts somehow more responsive to what dictates the logic, physics, and ontology, i.e., one must acknowledge the theoretical requirements coming from science itself for such development. However, this also involves stepping out of such disciplines and having a certain autonomy: since neither logic, science nor ontology were able to \textit{determine} the metaphysics of non-individuals, it seems that the only productive way to provide a metaphysical profile is the free development of a metaphysical profile for non-individuals, one that is tailored to fit quantum mechanics.

\citet{french2018jgps} argues that weak discernibility is a case in point. The argument is that \citet{saunders2003} used in a quantum-mechanical context the notion, originally developed by \citet{quine1976}, of weak discernibility to interpret quantum objects. Leaving aside the questions of whether this determines a metaphysical profile \citep[for a critical assessment of this claim, see][]{arenhart2017theo}, French's (\citeyear[p.~223]{french2018jgps}) point is this: ``metaphysics may still be useful to the philosopher of physics even if it is grounded in considerations that appear to have nothing to do with modern physics''. In other words, metaphysics, even if it is not directly focused on specific scientific questions, can contribute to science \textit{if applied in science}. This is made clear in passages such as the following:

\begin{quote}
    Quine himself was certainly not disinterested in science and although the notion of weak discernibility was clearly not motivated specifically by considerations of physics (quantum or classical), one could perhaps argue that the whole framework of his discussion of discernibility and identity may have been influenced by his reflections on science, even if only indirectly. \citep[p.~224]{french2018jgps}
\end{quote}

We wish, however, to stress that it is not enough to transport the metaphysical theory to the scientific context, or merely apply it in the scientific context so that it can provide a ``clear picture'', i.e. so that it can provide conditions to address Chakravartty's Challenge. A tailoring work must be done, that is, the metaphysical profile must be developed to obey the logical, scientific, and ontological characteristics of the theory in question. Of course, many times this metaphysical theory tailored to physics will employ a familiar vocabulary in its development (\citet[p.~227]{french2018jgps} recognizes this), for reasons of intelligibility. But the point we would like to emphasize is this: no metaphysical theory that floats totally free from physics can serve as a metaphysical profile for physics; the tailor's work is indispensable if we want to delve into the waters of `deep' scientific realism and respond to Chakravartty's Challenge.

Nevertheless, while this metaphysics is inspired by scientific theory, it is not directly related to it, as it will need philosophical elements unrelated to it. Now, suppose more than one metaphysical profile can be tailored to fit the physics. This is always a possibility, and it highlights the fact that no metaphysical notion comes from quantum theory itself. There seem to be no objective criteria to help us when it comes to theory choice \citep[see][]{benovsky2016, chakravartty2017}. 

Let us pause and reflect on what has been said so far. We argue that no metaphysics is guaranteed by anything other than metaphysics itself. So if we want to look for metaphysics, we shouldn't place our expectations of finding them in logic, physics, or ontology --- which is not to say that these disciplines are absolutely useless for metaphysical reflections \citep[see][]{arenhart2012, arroyoarenhart2019}, only that these disciplines are not capable of determining metaphysics \citep[see][]{arroyoarenhart2020}. This puts the discipline in a fragile position. For, if metaphysics reappears as a necessary condition for scientific realism (in its so-called `deep' version), its floating-free aspect raises a problem against the very possibility of scientific realism, which is metaphysical underdetermination.

Maybe things are not so bad. For example, we can quickly advance a metaphysical criterion coined by \citet[pp.~82--84]{benovsky2016} as `widen the net'. The idea is simple: we should not look at metaphysical theories, when applied to science, in isolation; instead, we should `widen our net' of investigation, and see how these theories relate not only to a scientific theory but to other \textit{neighboring} theories. Here comes Krause's (\citeyear{krause2005, krause2019}) suggestion: since physics is unable to decide between the metaphysics of individuality and non-individuality, perhaps \textit{chemistry} can do it. As \citet{krause2019} suggests, take the chemical reaction of methane combustion for the sake of an example: $CH_4+2O_2 \longrightarrow CO_2+2H_2O$. There is no way to tell whether the Hydrogen atoms in the methane molecule are \textit{the same} ones composing the water molecules: they are indiscernible. Other examples can be found in \citet{krause2005}, but the point made is: chemistry doesn't work without \textit{absolute indistinguishability}; thus we should prefer non-individuality over individuality for this (pragmatic) reason.

But which metaphysics of non-individuals should we adopt? There are several, or, as \citet[p.~1345]{arenhart2017synt} put it, the ```nonindividuals horn' produces its own underdetermination''. At this point, we are without the pragmatics on our side. Perhaps an alternative would stand out for being compatible with all disciplinary levels (logical, scientific, and ontological), which is precisely the tailor-made methodology for metaphysics that we had been proposing until then. But how to do it? Alas! This subject deserves the attention of an entire paper.\footnote{What is worst: notice that appealing to indiscernibility in chemistry does not exclude the possibility of using some of the principles of \textit{individuality} in the Transcendental Individuality family, which would just show that underdetermination is left precisely on the same point!}

\section{Conclusion}
\label{arenhart-arroyo-sec:6}

Let us take stock and check what has been achieved. The plan, recall, was to find some possible source for the nature of quantum objects as non-individuals, and that with some kind of good epistemic authority, which could in some sense be justified. Certainly, the idea of non-individuality arises with quantum mechanics, in its early days, as the founding fathers fought to grasp what kind of thing was it that the theory was attempting to describe; but in the face of the current multiplication of alternative approaches, many of them quite conservative in metaphysical terms, one may ask for the epistemic credentials of the view. After all, being the first to enter the stage in the metaphysics of quantum mechanics does not confer to it any kind of epistemic privilege. So, are we justified in believing that quantum entities are non-individuals, in any given sense? This may be asked both by realists, attempting to answer to Chakravartty's Challenge, as well as by anti-realists willing to check how the world could look like according to a specific reading of the theory; the latter could face it as an exploratory investigation, perhaps even to show underdetermination and throw it back on the realist. We have supposed that there is a legitimate interest in discovering some kind of connection between the idea of non-individuality and some privileged source that could help us shed some light on the notion.

In this paper, we explored some of the options that are typically mixed in the literature as a source for the non-individuals character of quantum entities. Logic, physics, ontology, and metaphysics proper. All of them are found wanting, or, in other terms, not responsible for the metaphysical character or nature of non-individuals attributed to quantum objects. Logic does not allow us to confer the required respectability on non-individuality, rather, it requires that such a notion be given beforehand. Physics, it is well known, cannot be held responsible for quantum non-individuality; indeed, there is nothing in physics that answers for metaphysics. The metaphysics must be put there from somewhere else, it seems. Ontology is not the source also, given the fact that an ontology of objects is not even the only possibility for quantum theory. It seems that disputes in ontology all seem to declare that their posits are a result of the needs of physics, but then, physics is certainly not sure of what it needs in ontological terms, anyway. Distinct options seem all to be vindicated by the theory, but on a closer view, what is obtained is at best a kind of compatibility with the theory. Finally, metaphysics itself cannot be responsible for non-individuality. Besides the fact that individuality is also compatible with physics, there are just too many ways to understand what non-individuals are, all of them equally good for physics. One looking for metaphysical tools in the field of traditional metaphysics just does not know which tool is the appropriate one, and metaphysicians can do no better than to open the gigantic toolbox and offer their best smile.

The idea that quantum entities are individuals or not is clearly not justified by any kind of source that finds itself already involved in the debate. In this sense, we generalized the argument of metaphysical underdetermination: it is not just physics that does not determine its metaphysics. Logic does not determine its metaphysics nor ontology, neither ontology do so. In this sense, we should not place our hopes to justify metaphysics with the help of other areas of knowledge --- recall that this is precisely the naturalist project --- because the epistemic warrant is not something that metaphysics can inherit. Metaphysics as a discipline finds itself floating free not just from physics but pretty much from everything else. This is not a problem specific to non-individuality. The epistemic authority to be attributed to metaphysical claims is just as difficult to find for any other metaphysical theory that wants itself attached to a physical theory. Clearly, the source of the problem is to be found on the very idea of conferring \textit{epistemic} authority to a metaphysical notion, in its relation to quantum mechanics. Perhaps we have been wrong in looking for some kind of privileged authority that could be attributed to metaphysical claims. Or, perhaps, we have been wrong in looking for a metaphysical complement to science that can still be somehow justified. But addressing these issues is more than we could do now. 

\printbibliography[heading=subbibliography]
\end{refsection}

\end{document}